# A quantum model for Johnson noise

José-Ignacio Izpura

***Abstract*— Johnson noise is a small random voltage that appears between terminals of any resistor interacting with its thermal bath at temperature T. It looks like continuous, but the discreteness of the electrical charge suggests its discrete origin coming from the charge noise due to random translocations of individual electrons between terminals. The capacitance allowing these translocations would quantize the energy entering the resistor in this way, thus acting as the "antenna" of the resistor to pick up thermal energy in the form of charge unbalances (fluctuations of energy) between its terminals. The subsequent relaxations of these fluctuations by the conductance G=1/R of the resistor (the collective reaction of all its carriers) would give rise to its Johnson noise. This collective reaction to dissipate fluctuations of energy caused by individual electrons, agrees with the Fluctuation-Dissipation framework that Callen and Welton proposed in 1951 for noisy processes.**

## I. INTRODUCTION

Although the quantum proposal for noisy processes [1] appeared in 1951, a quantum model for electrical noise (EN) still is lacking. I mean a model showing the type of fluctuations of energy generating Johnson noise in resistors, not abstract ones invoking [1] as a dogma to grant the existence of this noise in devices represented by circuits without reactive elements giving room for fluctuations of energy. Measuring EN in resistors one meets their low-pass, Lorentzian spectrum of Johnson noise proportional to their resistance R. Its amplitude at low frequency $S_V$=4kTR $V^2$/Hz, where T is temperature and k the Boltzmann constant, drops as frequency $f \rightarrow f_c$=G/(2πC), its cutoff frequency defined by their conductance G=1/R and their capacitance C in parallel. This spectrum appears from thermal equipartition (TEQ) in this relaxation cell [2].

Regarding C, we took initially $C_{mat}$ from the permittivity $\varepsilon=\varepsilon_r\varepsilon_0$ of the material between its two equipotential terminals (plates) at distance $d$ that Fig. 1 shows by a 1-D model for this two-terminal device (2TD) whose plates allow to apply (and sense) electric fields along $d$ in this device. Due to their capacitive coupling, the current entering this 2TD by one plate is exactly equal to that collected by the other. Hence, resistors offering a pure resistance R between terminals do not exist. Existing ones offer their resistance R with some capacitance $C_{mat}$ in parallel due to the resistivity ρ and permittivity ε of their inner materials.

The conductance G=1/R of each resistor entails its ability to dissipate (to convert into heat) electrical energy present in $C_{mat}$ that fluctuates with time. This was the basis of my first model for EN agreeing with [1] that we used in [2-3] and recently in [4] to consider how resistors sense their thermal bath. In this model, resistors undergo fluctuations of capacitive energy that their conductance dissipates subsequently.

While reading [1] in 2008, I saw that my action/reaction model for Johnson noise based on signal theory, could fit this quantum framework. To show what I mean, let me design a 50Ω resistor to match a lossless transmission line (TL) of $Z_0$=50Ω (line impedance) from material of ρ=5 Ω×cm like silicon doped by $N_d \approx 10^{15}$ $cm^{-3}$ donors. Alloying metal on two faces of a cube of this Si to have two terminals at distance $d$=1mm=L (its side length) it would show R=ρ×$d$/$A_P$=ρ/L=50Ω between them. Neglecting surface effects and assuming full donor ionization at room T for this low $N_d$, the volume $V_Q=10^{-3}$ $cm^3$ of this 2TD would contain n=$10^{12}$ carriers (free electrons in the conduction band of its Si material). Fig. 1 sketches this Si device.

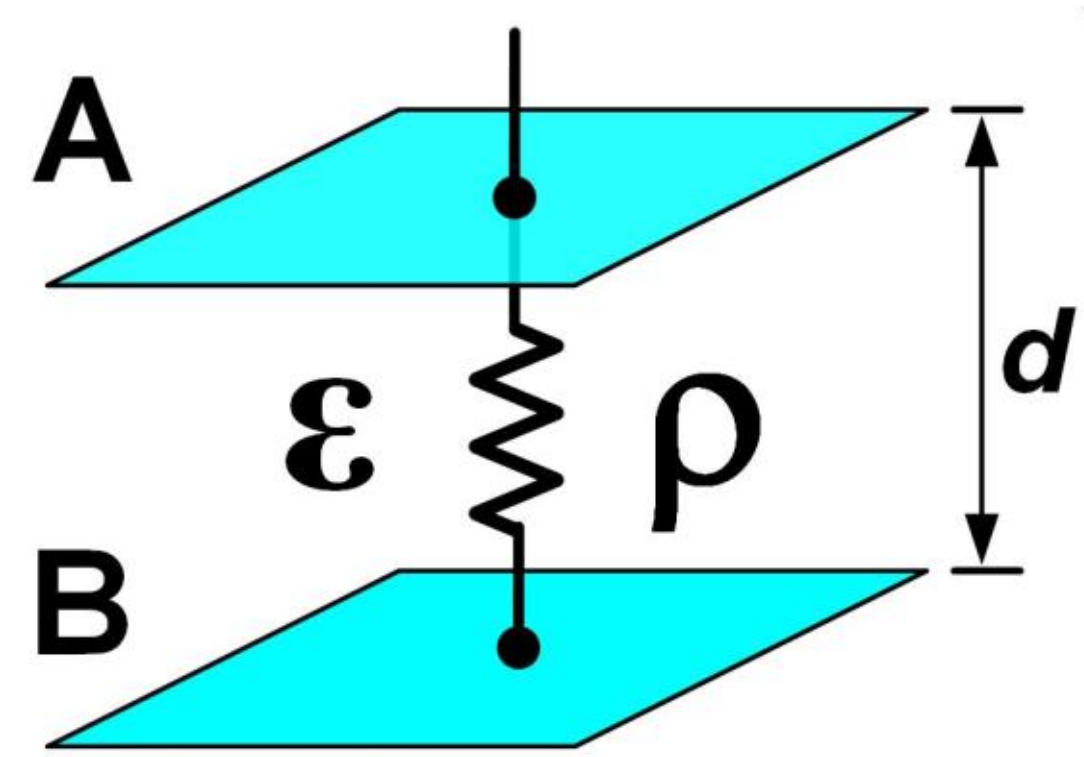

**Fig. 1.** Basic geometry of a two-terminal device (2TD) like a resistor (or capacitor) highlighting the capacitive coupling that exists between its terminals (plates of area $A_P$) at distance d.

From the permittivity of silicon $\varepsilon \approx 12\varepsilon_0$, the capacitance of my resistor would be $C_{mat}$=ε×$A_P$/$d$≈0,1pF. Therefore, the time constant R×$C_{mat}$=5 ps of this device is the dielectric relaxation time $\tau_d$=ε×ρ of its material. A sudden voltage created between its terminals would decay with $\tau_d$=5 ps. Hence, my device is a relaxation cell of $\tau_d$=5ps, whose $f_c$=1/(2πR$C_{mat}$)=32GHz shows its limitation to match my TL, as I warned in [3]. At $f$=32 GHz, its R=50Ω shunted by the capacitive impedance of $C_{mat}$ gives $Z_L$=25-j25Ω ending my TL. For $P_{in}$ watts coming from the TL at 32GHz, only 0.8$P_{in}$ would enter my resistor and $P_{in}$/5 would be reflected to the TL. This frequency-induced mismatch makes the matching of my cubic resistor acceptable only up to $f$≈6.4 GHz, where it only would reflect $P_{in}$/100.

Work supported in part by the Spanish Ministerio de Ciencia y Tecnología under the PID2020-118410RB-C22 project and by the European Union under the ESPRIT 25112 and 35112 projects. *(Corresponding author: J. I. Izpura).*

José Ignacio Izpura is a Full Professor of the Universidad Politécnica de Madrid (UPM), Madrid, Spain, in its Department of Aerospatial Systems, Air Transport and Airports of the ETSIAE. (e-mail: joseignacio.izpura@ upm.es).

Applying TEQ to its $C_{mat}$, the mean square voltage of its Johnson noise is $<(v_n(t))^2>=kT/C_{mat}$ $V^2$. This voltage driving its R=50Ω means that my cubic device in thermal equilibrium (TE) at temperature T would dissipate $P_{dis}=<(v_n(t))^2>/R=kT/\tau_d$ watts. Hence, the material of my resistor defines its dissipation in TE no matter its volume or size. Two cubic resistors like mine, put in series would give a 100Ω resistor that would dissipate the same power than mine of 50Ω at the same T. $P_{dis}$=828 pW is the mean power $P_{abs}$ that should enter the volume $V_Q$ of my 2TD.

This mean power $P_{abs}$=828 pW that my device would take from its thermal bath, would enter it by sudden translocations of single electrons between its terminals causing instantaneous fluctuations of energy in its $C_{mat}$ [4]. Assuming that the energy $U_f=q^2/(2C_{mat})$ of each fluctuation is removed from $C_{mat}$ before a next fluctuation occurs (on average, see later), the mean rate λ of these fluctuations should be [2-4]:

$$\lambda \times U_f = P_{dis} \Rightarrow \frac{\lambda q^2}{2C_{mat}} = \frac{kT}{RC_{mat}} \Rightarrow \lambda = \frac{2kT}{q^2 R} \Rightarrow$$

$$G = \frac{1}{R} = \lambda \times \frac{q^2}{2kT} = \lambda \times \frac{q}{2\left(\frac{kT}{q}\right)} = \lambda \times \frac{C_f}{2} \qquad (1)$$

This rate λ not depending on $C_{mat}$ deserves attention. Using "slower material" of higher $\tau_d$ would increase $C_{mat}$ by the same factor β that would reduce $U_f$. Keeping λ would decrease by β both the power $P_{abs}$, and the new $P_{dis}$ of its Lorentzian spectrum of Johnson noise whose $f_c$ would be β times lower. One would say that the rate λ not depending of $C_{mat}$ seems to hold for the "capacitive antenna" that resistors would use to pick up thermal energy. Using "fast" materials of low $\tau_d$ to grant the dissipation of the energy $U_f$ before the next fluctuation occurs Eq. (1) holds. For materials like GaAs or even Si with $N_d>10^{18}$ $cm^{-3}$ giving $\tau_d$ in the fs range, thus $f_c$>1 THz, this is not a problem.

By contrast, using "slow material" so that its $\tau_d$ exceeds the mean time between fluctuations $T_{avg}=1/\lambda$, the decaying pulses of voltage caused by the fluctuations would overlap in time. This piling-up of voltage in $C_{mat}$ (that speeds the removal of its energy) can be useful for noise in capacitors, but falls out of the scope of this paper. Regarding the fast dissipation of $U_f$, the last form of Eq. (1) for G=1/R being proportional to the rate λ, came from this worry. From this result, the rate λ for the absorption of thermal energy by packets of $U_f$ joules each in the resistor appears coupled with its dissipation. Reading carefully Eq. (1): "*The conductance G=1/R of a resistor seems to come from a random series of chances to dissipate packets of energy at the same rate λ of its chances to absorb packets of thermal energy*".

Experiments show that when its heating effect is low, the dc current biasing a resistor does not vary its Johnson noise. From Eq. (1) for $G=\lambda\times(C_f/2)$ and mostly from its tiny capacitance $C_f=q/V_T$, Section IV proposes a disruptive form of dissipative current in resistors that would not affect their Johnson noise. Now, let me leave shortly aside dissipation to consider how fluctuations create the voltage called Johnson noise. About this random voltage, let me consider in my cubic resistor the voltage step due to the translocation of $q=1.6\times10^{-19}$ C between its terminals. This gives a step of $\Delta V=q/C_{mat}$=1.6 μV with null risetime due to the simultaneity of the currents that enter and leave my resistor by its terminals at distance $d$.

Taking dc current as a flow of positive charges, this current going down in Fig. 1 would entail electrons going up from plate B (source) to plate A (drain), across $V_Q=A_P\times d$ (the volume of this cubic resistor between plates of area $A_P$). The charge $-q$ C of an electron suddenly arrived in plate A at the instant $t_0$ from $V_Q$, would set an electric field $E=-q/(\varepsilon A_P)$ "pricking" plate A. This field crossing $V_Q$ at the speed of the electromagnetic wave, is a synchronizing signal to inform plate B that the charge $+q$ C that it acquires *at the instant* $t_0$, causes on it the field $E=q/(\varepsilon A_P)$ leaving plate B to "prick" plate A at $t_0$. Since the $-q$ C (electron) arriving in plate A, and $+q$ C appearing in plate B (as if an electron of charge $-q$ had left it) are events that a voltmeter would take as simultaneous at $t=t_0$, I considered that a single electron can pass instantaneously between terminals of a 2TD [4] by their capacitive coupling. Hence, displacement currents of null dwell-time and weight $q$, translocating single electrons between terminals, would produce instantaneous fluctuations of energy in the $C_{mat}$ of my cubic resistor.

Each impulsive current would set a voltage ΔV=1.6 μV in my 2TD or a fluctuation of $U_f=q^2/(2C_{mat})$ joules in its $C_{mat}$. An electron of its terminal A absorbing $U_f$ would appear in its terminal B leaving a charge $+q$ in the former. The low value $U_f$=0,8 μeV for my cubic resistor (31 ppm the thermal energy kT at room T) suggests that these translocations should be very frequent in my device, whose impulse response h($t$) is a pulse $h(t)=h_0\times\exp(-t/\tau_d)$ with $h_0$=1.6 μV decaying with $\tau_d$=5 ps. Since the Fourier transform of h($t$) is a Lorentzian of $f_c$=32GHz, the λ responses h($t$) per second appearing on average and randomly in my resistor, would give its Lorentzian spectrum of Johnson noise with $f_c$=32GHz (Carson's Theorem).

For this to be so, its fluctuations should take place at random times and with random sign (50% positive and 50% negative on average). Given their null dwell-time, nothing prevents this total randomness. This model based on signal theory that I have been using up to now, is the basis of the new quantum model for EN that I will complete along this work. For this task, let me consider my 2TD as two metal plates cladding a silicon cube containing $10^{12}$ "carriers" ready to sense any field along $d$ (E=V/$d$) coming from the voltage V existing between its two plates or terminals. Added to it, I will consider this familiar, but puzzling result for a resistor: that a bias current $I_C$=V/R does not add voltage noise to its Johnson noise (provided $I_C$ does not rise noticeably its T). Reasons trying to explain this lack of shot noise assigned to bias currents appear in the literature. In this work, however, I will give this disrupting one:

*"In macroscopic resistors, shot noise assigned to their bias currents is not observed because these currents are dissipative currents, not displacement currents deserving the assignment of shot noise like those that generate their Johnson noise."*

This is why Section II considers the two types of electrical current that exist in a resistor while its Johnson noise is being generated from random voltage responses h($t$) that relax with time. It also shows that assigning shot noise to dc currents in resistors is unfounded because the shot noise of displacement currents that they have, explain perfectly their Johnson noise. Section III gives empirical support to these findings. Sections IV, V and VI complete the quantum model for EN and the new one for dissipative current that we propose in this work.

## II. Shot noise in capacitors seems elusive in resistors: two "very different" currents

A resistor under open circuit conditions shows its Johnson noise while its net current is null. The circuit that Fig. 1 suggests is a capacitor and a resistor that exist in the volume $V_Q$ of my resistor. In this lumped model, one can separate the capacitive current of $C_{mat}$ and the dissipative one of R to say that current "leaving $C_{mat}$" by its upper terminal "enters R" by its upper terminal, thus nulling the net current across this RC cell. If a voltage $h_0$ started to relax in it, the voltage $V_{AB}$ of Fig. 1 would be $h(t)=h_0\times\exp(-t/\tau_d)$, thus decaying with $\tau_d=R\times C_{mat}$ as its time constant. For $h_0>0$ (i. e. for $V_A>V_B$), the current $i_c(t)=h(t)/R$, would go down in Fig. 1. It would be the "dissipative current of R coming from the displacement current $i_d(t)=C_{mat}\times(dh(t)/dt)$ going up in $C_{mat}$", whose voltage $h(t)>0$ keeps a negative time-derivative as it relaxes under this sum $i_c(t)+i_d(t)=0$.

This null sum is the differential equation that gives rise to $h(t)=(q/C_{mat})\exp(-t/\tau_d)$ as a natural frequency of the conversion gain $V_0/I_i$ of the resistor. Thus, h($t$) is voltage $V_0(t)$ that can exist while $I_i=0$ (e. g. in TE) because this relaxing voltage remains in $C_{mat}$ after its excitation by a current already gone. Since the noise voltage of a resistor comes from its $\lambda$ responses h($t$) per second caused by its fluctuations, to know the noise voltage of my resistor of Fig. 1, let me consider one of the pairs of events (fluctuation-dissipation) that would produce its Johnson noise.

The first event is the displacement of a single electron from terminal A to terminal B that sets a voltage $V_{AB}=h_0>0$ in this device. Due to $C_{mat}$, the energy $U_f$ that the electron of terminal A must absorb to appear in terminal B is low and it undergoes this translocation instantly. The effect that we can measure of this impulsive current is a step h($t$) of $\Delta V_{AB}=q/C_{mat}$ volts that starts to decay with time constant $\tau_d$. It decays because $h(t)\neq 0$ in $C_{mat}$ drives $i_c(t)=h(t)/R$, a conduction current that converts its energy into heat left in the material. To mean something that $i_c(t)$ does, its name will be "dissipative current" hereafter. Given that terminal A of this resistor is at distance $d$ over its terminal B in Fig. 1, its current $i_c(t)$ for $V_{AB}=h(t)>0$ goes down.

Since the time derivative of $V_{AB}>0$ is negative, the current $i_d(t)=C_{mat}\times(dh(t)/dt)$ goes up in Fig. 1 and cancels at each instant of time the $i_c(t)$ going down in $V_Q$. Each $i_d(t)$ displacing charge in space is a pulse decaying with time constant $\tau_d$, whose time integral is $q$. It would bring back "slowly" the electron that the fluctuation displaced instantly in opposed sense. Thus, each $i_d(t)$ discharges $C_{mat}$, a job often assigned to $i_c(t)=h(t)/R$, whose time integral also is $q$ [4]. However, if $i_d(t)$ already discharges $C_{mat}$, its accompanying $i_c(t)$ *should not do it*. It should dissipate energy without displacing charge to be "orthogonal" to its mate $i_d(t)$ that displaces charge without dissipating energy.

From the noise viewpoint, Johnson noise comes from these two pulsed currents of opposed sign cancelling mutually at each instant of time. This exact cancelation does not give the noise of two antagonistic flows of carriers that cancel one to each other *on average*. Think of the saturation currents of a junction diode in TE generating its voltage of noise that PSPICE could not simulate [2]. While charge unbalances relax in a resistor, it has pulses $i_d(t)$ of displacement current entailing shot noise due to the charge they "move" between terminals. Because $q$ is the time integral of each pulse $i_d(t)$, multiplying $q$ by the rate of pulses $\lambda$, the total displacement current in the resistor is: $I_{Tot}=\lambda q$ amps. The shot noise density of these random pulses of current as $f\rightarrow 0$ will be $S_{Ishot}=2qI_{Tot}$, thus giving $S_{Ishot}=2\lambda q^2$ $A^2$/Hz as the amplitude of its Lorentzian spectrum coming from pulses of displacement current decaying with $\tau_d$. From Eq. (1) one finds $S_{Ishot}=2q^2\times(2kT)/(Rq^2)=4kT/R$ $A^2$/Hz that is the familiar density of noise called Nyquist noise [5].

If the displacement currents $i_d(t)$ of a resistor in TE already account for all its Johnson noise, its dissipative currents $i_c(t)$ should not add noise. Hence, assigning shot noise to dissipative current is unfounded, as experiments show over and over. In regard to the spectrum of this Nyquist noise, it should be a Lorentzian with $f_c=1/(2\pi\tau_d)$ as its cutoff frequency coming from each current $i_d(t)$ that gives rise to $S_{Ishot}$. In other models for EN, it could be flat up to the limit $f_N=kT/h$ of [5] (≈6THz at room T) but in my quantum model, its cutoff frequency is $f_c=1/(2\pi\tau_d)$. The 4kT/R $A^2$/Hz of my cubic resistor, coming from its random pulses $i_d(t)$ decaying with $\tau_d=5$ ps at room T, only would be flat up to $f\approx 6$ GHz. The $f_N\approx 6$ THz [5] that could have to do with its impulsive fluctuations, falls out of this work.

I will leave aside this interesting subject on the measurement of electrical current to exploit the new results of my first model for EN. I mean the noiseless nature it predicts for the dissipative currents that cancel the displacement ones that produce Johnson noise. Regarding the pulses $i_c(t)$, they form a current that most people would take as charges drifting between terminals, whose shot noise is not observed. Note that $i_d(t)$ and $i_c(t)$ are pulses of amplitude $q/\tau_d$ amps decaying with $\tau_d$, thus pulses with the same Fourier transform. If $i_c(t)$ entailed any charge being displaced along $d$, it would add shot noise increasing the Johnson noise. This would contradict the Nyquist noise that we infer from the Johnson noise that we measure in resistors.

Therefore, I will not assign shot noise to dissipative currents knowing that for people taking any electrical current as discrete charges "flowing" between terminals, this null assignment of shot noise is unbelievable. Since my first model for EN that agrees with [1] predicts dissipative currents being noiseless, let me show my own empirical evidence on this lack of shot noise assigned to a current biasing a resistor.

## III. Evidence of noisy displacement currents coexisting with noiseless dissipative currents

Textbooks assuming tacitly a current $I_C$ as a flow of discrete electrons give $S_{Ish}=2qI_C$ $A^2$/Hz as its density of shot noise at low frequencies (as $f\rightarrow 0$). Using this formula, one accepts $S_{Ish}$ and its notion of travelling charges, which is true for electrons that travel in space-charge regions or in vacuum [2, 3]. However, to contend that these travelers exist, we must find empirically their $S_{Ish}$. This entails to convert $S_{Ish}$ into voltage density $S_{Vsh}$ (effect) that once measured by a voltmeter, allows inferring the current density $S_{Ish}$ that has been its cause. This conversion and this inference are required tasks to measure electrical current. For a resistor, its own resistance would convert $S_{Ish}=2qI_C$ A$^2$/Hz into $S_{Vsh}=2qI_CR^2$ V$^2$/Hz at low frequencies.

Now, let me use the circuit of Fig. 2 to show the lack of shot noise for a dc current $I_C\approx 0,7\mu$A biasing its three resistors at the input of its low noise amplifier (LNA). Joining the two resistors of 3MΩ by a jumper, we form a resistor of $R_t\approx 1$MΩ (1,2MΩ shunted by 6MΩ) whose Johnson noise is measured at room T. This balanced $R_t$ exploits the differential amplification of our LNA EGG-PAR113. A 5V battery shunted by 1µF (to grant its low impedance for $f$>10Hz) allows to set $I_C$=5V/7,2MΩ≈0,7µA in these resistors. Replacing the jumper by the battery, the same $I_C$ is set in each resistor. The two resistors of 3MΩ in series give a Nyquist density $S_{I2}$=4kT/6MΩ=2,8×10$^{-27}$ A$^2$/Hz whereas the shot noise assigned to their current $I_C$ is $S_{Ish2}=2qI_C=2,25\times 10^{-25}$ A$^2$/Hz ($S_{Ish2}\approx 80$ times $S_{I2}$).

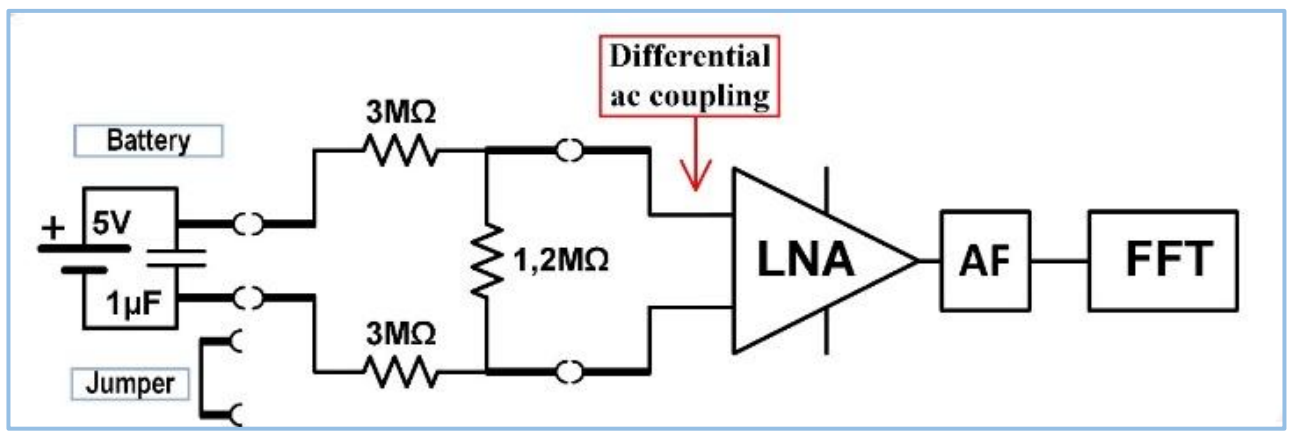

**Fig. 2.** Setup to show that a dc current $I_C\approx 0,7\mu$A does not increase the Johnson noise of an $R_t$=1MΩ resistor at room T. Note that $R_t$ and the input $C_{stray}$ of this LNA adds a third low-pass filter to the two antialiasing ones (AF) of our FFT analyzer sampling at 250 ksamples/second (see the text).

The shot noise $S_{Ish1}=2qI_C=2,25\times 10^{-25}$ A$^2$/Hz assigned to $I_C$ in the resistor of R=1,2MΩ, is sixteen times its Nyquist noise $S_{I1}$=4kT/R=1,4×10$^{-26}$ A$^2$/Hz, thus $S_{Ish1}=16S_{I1}$. The total noise density shunting $R_t$=1MΩ will be $S_I=4kT/R_t=1,66\times 10^{-26}$ A$^2$/Hz (its own Nyquist noise) plus the shot noise assigned to $I_C$. From $I_C$ in 3+3=6 MΩ we have $S_{Ish2}=2qI_C=2,25\times 10^{-25}$ A$^2$/Hz plus an equal amount $S_{Ish1}$ assigned to $I_C$ in 1,2 MΩ. The total shot noise assigned to $I_C$ ($S_{Ish}=2\times 2qI_C$) is $S_{Ish}=27S_I$. Hence, the shot noise assigned to $I_C$ should rise the Johnson noise of $R_t$ by 14,5 dB (28 times). Fig. 3, however, confirms one more time that we should not assign shot noise to $I_C$. This comes from the noise voltage densities (V$^2$/Hz) that it shows at the output of the LNA, thus those at its input multiplied by the fixed gain (10$^4$ times, 80 dB) used in these measurements.

Since the direct measurement of electrical current is not possible, we have converted densities of current into densities of voltage (V$^2$/Hz) that our spectrum analyzer (a voltmeter) can measure. Although I have used densities of noise current in A$^2$/Hz, it is because all of them add in parallel. The red graph **b)** of Fig. 3 is 10$^4$ times the voltage noise density of the $R_t$=1MΩ resistor of Fig. 2 with the jumper set, thus 10$^4$ times the Johnson noise $S_V=4kTR_t$ at room T. This noise that is $S_V=1,66\times 10^{-14}$ V$^2$/Hz (i. e. -137.8 dB) plus the 80dB gain of our LNA should give a flat region at ≈-57.8 dB at low $f$. This is the flat region of curve **b)** lying at -57,5 dB. Although I could trim the gain of this old LNA properly, I have used its fixed gain "as it is today", because the proof I am giving does not need such accuracy.

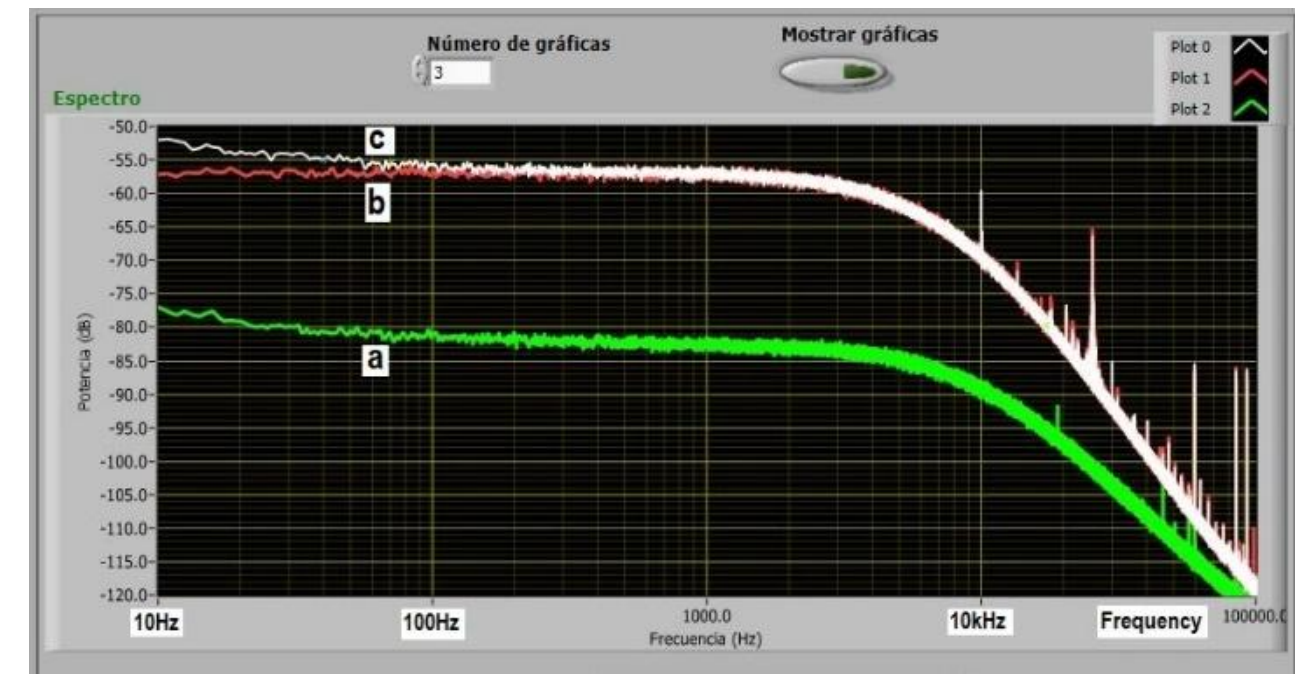

**Fig. 3.** Noise voltage found in the circuit of Fig. 2 in these cases: a) with the inputs of the LNA shorted together; b) Noise voltage of the $R_t$=1MΩ resistor at the input of this LNA with the jumper connected ($I_C$=0); c) Noise voltage at the input of this LNA when the battery replaces the jumper ($I_C\approx 0,7\mu$A).

Curve **c)** of Fig. 3 is the density $S_V$ of this resistor biased by $I_C$=0,7µA when the battery replaces the jumper, where we can see that graphs **b)** and **c)** overlap in the 100Hz-100kHz band. Thus, $I_C$=0,7µA does not change the Johnson noise of my 1MΩ resistor. Its density of noise voltage for $I_C\approx 0,7\mu$A is $S_{Vtot}\approx S_V$, not $S_{Vtot}=28S_V$ (14,5 dB higher) as one expects from assigning shot noise to $I_C$. The $S_{Vtot}\approx S_V$ that I have measured means that the bias current $I_C$ does not affect the "*agitation of charge*" that produces its Johnson noise, following the titles of [5, 6]. Since devices with displacement currents clearly show $S_{Vsh}>0$ [2], I contend that displacement currents give shot noise that other currents like dissipative ones do not give, thus agreeing with the predictions of my first model for EN.

Hence, shot noise from bias currents should not appear in resistors and its lack for $I_C\approx 0,7\mu$A in our resistor, no longer is the puzzling result it was in the past. It is the proof of its null noise that my model predicts. Concerning $P_{Dis}$=(5V)$^2$/7,2MΩ (the power of 3,5µW that $I_C$ dissipates in the resistors of Fig. 2) its heating effects seem negligible for these devices whose size is shown in Fig. 4. This explains why its noise voltage densities for $I_C$=0,7µA and for $I_C$=0 overlap within experimental error down to $f$=100Hz, a frequency where $I_C$ reveals that these resistors are not free from *resistance noise* giving rise to the excess noise [3, 7] that appear for $f$<100Hz in Fig. 3.

Note that the capacitance $C_{in}/2$=7,5 pF of the balanced input of our LNA used in differential mode shunts $R_t$=1MΩ in Fig. 2. This $C_{in}/2$ comes from its two inputs ($C_{in}$=15 pF each) that are in series while the noise voltage of $R_t$=1MΩ is amplified in differential mode. This capacitance must include the wiring one of the two coaxial cables that connect our resistor to each input of the LNA, see Fig. 4. Given the length (≈14cm) of each cable, its $C_{coax}\approx$16pF, would add 8pF to $C_{in}/2$. Hence, the resistance $R_t$=1MΩ has $C_t$≈15,5pF in parallel that forms a first order, low-pass filter of $f_{C3}$=1/(2π$R_tC_t$). This is the third AF mentioned in the footnote of Fig. 2, whose cutoff frequency $f_{C3}$≈9,9 kHz sets the power $P_{TE}$=kT/($R_tC_t$) that our 1MΩ resistor dissipates in TE at room T. With three low-pass AF of $f_C$≈10 kHz each, aliasing effects under 10 kHz are negligible in these data taken at the sampling rate of 250.000 samples/second.

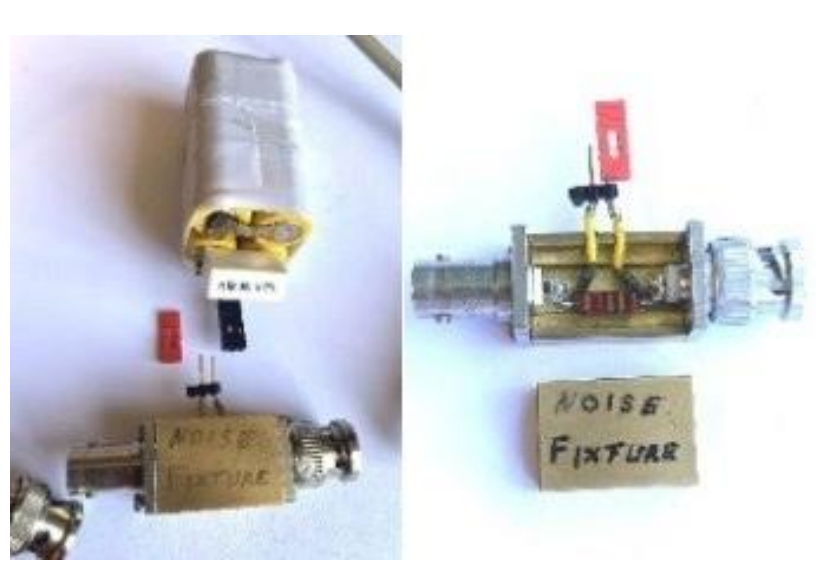

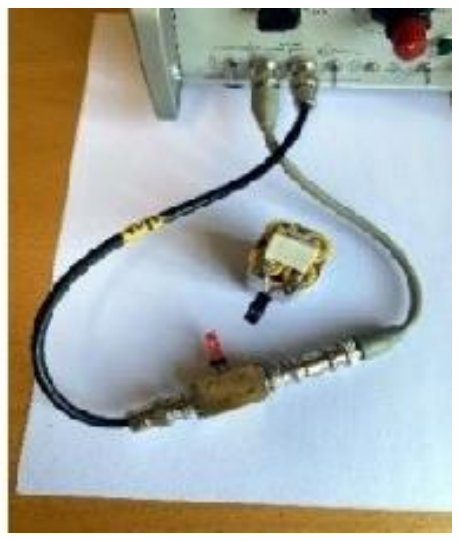

**Fig. 4** Noise fixture showing its two pins where the battery that sets $I_C$ replaces the jumper joining them for $I_C$=0. See the two coaxial cables of similar length from this test fixture to the inputs of the EGG-PAR 113 Low Noise Amplifier.

Comparing the dissipation of $I_C$ in the resistors of Fig. 2 $P_{Dis}$≈3,5µW and their dissipation in TE $P_{TE}$=2,6×$10^{-16}$ W, Fig. 3 becomes *spectacular*. The overlap of its **b)** and **c)** graphs shows that $I_C$≈0,7µA increases their dissipation 1,3×$10^{10}$ times (i. e. 101dB) without disturbing its Johnson noise generated from a fluctuation-*dissipation* dynamic. This huge factor suggests that current dissipating energy must be totally different from charge that moving between terminals (e. g. drifting electrons) would affect the Johnson noise. This conviction led to the disruptive model of the next Section.

## IV. DISSIPATIVE CURRENT FREE FROM SHOT NOISE

Let me resume the subject of dissipation that I left aside in the second paragraph below Eq. (1) by this sentence about it: "*The conductance G=1/R of a resistor seems to come from a random series of chances to dissipate packets of energy at the same rate λ of its chances to absorb packets of thermal energy.*

Given a resistor, multiplying its conductance G=λ($C_f$/2) of Eq. (1) by the square of its voltage V between terminals gives the power $P_{Dis}$=R×$(I_C)^2$ dissipated in it by its current $I_C$. Hence, charging $C_f$ with V volts λ times per second, would take an electrical power equal to $P_{Dis}$. The huge values of λ and the tiny one of $C_f$, both suggest that $C_f$ has to do with each carrier in the resistor. Going to my cubic one of R=50Ω, Eq. (1) states that it would undergo λ=6,4×$10^{15}$ dissipations per second at room T. This rate could be reached by its n=$10^{12}$ carriers, dissipating many times per second the energy $U_f$=($C_f$/2)×$V^2$ that V would load in $C_f$. Regarding $C_f$=$q$/$V_T$ (6.2×$10^{-18}$ F at room T), I found it during my Ph. D. Thesis on DX centers in AlGaAs. It would be a kind of "*capacitive trace of each carrier"* trapped between the terminals of a resistor, as I will show.

To study the mean power dissipated by $I_C$ in the resistor, let me take $C_f$ sensing (and thus, loaded with) the voltage V=R×$I_C$ between terminals of a resistor biased by $I_C$. From Joule's Law for $P_{Dis}$, Ohm's Law (V=R×$I_C$) and Eq. (1) for G, this power is:

$$P_{Dis} = R \times I_C^2 = \frac{(R\times I_C)^2}{R} = G \times V^2 = \lambda \frac{1}{2} C_f V^2 = \frac{V^2}{R} \qquad (2)$$

Eq. (2) shows that discharging λ times per second the $C_f$ thus charged would waste an electrical power equal to the power that $I_C$ dissipates in this 2TD. People aware of switched capacitor circuits dissipating electrical power without resistors will notice this "capacitive" form to dissipate electrical energy. Note that biasing the resistor by $I_C$ allows voltage changes like Johnson noise to exist between terminals. Biasing it by a dc source of V volts would "bury" this noise in its fixed voltage V. Eq. (2) that accounts for the enhanced dissipation $P_{Dis}$>>$P_{TE}$ making Fig. 3 spectacular, also accounts for $P_{TE}$ because V=0 (null mean) is the dc term of the Johnson noise of mean square kT/$C_{mat}$ $V^2$ in TE that led to Eq. (1). To explain in more detail the disrupting notion of $C_f$, let me say that this tiny capacitance sensing the voltage V, would store this energy:

$$U_{joul} = \frac{1}{2} C_f \times V^2 = \frac{1}{2}\frac{q}{V_T} \times V^2 = \frac{q^2}{2kT} \times V^2 \qquad (3)$$

Dividing Eq. (2) by Eq. (3) gives the λ packets (of $U_{joul}$ each) that the conductance G=1/R dissipates each second. This notion on G as an ability to dissipate λ packets of energy $E_{joul}$ each second suggests a dissipation process able to keep the Johnson noise of a resistor while a bias current $I_C$ increases by orders of magnitude its dissipation of TE. This process is one releasing to the lattice a packet of energy $U_{joul}$ each time a fluctuation occurs. This worry has brought back the tiny $C_f$ that I met years ago, studying donor atoms in AlGaAs to have "free electrons" or "carriers" in the conduction band (CB) of this material. Used to donor atoms of Si in AlGaAs, let me consider one of them as the *spherical dipole* that form its cation $Si^+$ of +$q$ C, screened by its cloud of charge -$q$ C (outer electron).

This dipole exists when this electron is "trapped" by the $Si^+$ cation as a shell of charge -$q$ at mean distance $d_0$ that could be its "radius". This electron does not contribute to the G=1/R of a resistor of macroscopic size because the force it senses due to the field V/$d$ (proportional to V×$d_0$/$d$) is low for reasonable V values in devices with terminals separated by $d$>>$d_0$. Although this force would deform this dipole, it would not break it and the same would hold for the fields of its thermal bath at low T. This dipole, However, could store thermal energy by its mean $d_0$ at each T varying as $d_0$(T). For fixed charges +$q$ and -$q$ in its "plates", their mutual capacitance C(T) would decrease as $d_0$(T) increases with T. In this way the energy $U_{sph}$(T)=$q^2$/[2C(T)] that it would store in the volume $V_{sph}$ enclosed by its outer electron, would increase with T. This expansion of $V_{sph}$ continues up to a temperature where this cloud of charge would expand to occupy suddenly the whole volume $V_Q$ of the device.

In this situation the donor atom is said "ionized" because its outer electron no longer is close to it. Its new distribution in $V_Q$ requires Bloch functions in its wavefunction to consider the periodicity of the atomic lattice. Hence, this electron that seems free, actually is in a bigger trap of volume $V_Q$, between the two terminals of this device. However, this cloud of density $-q/V_Q$ C/cm$^3$ in $V_Q$ can sense the voltage V between terminals and can react to any field E=V/*d* set in $V_Q$. To know what this reaction is, I need *a reason for its uniform distribution* in $V_Q$. Since it is a movable cloud of negative charge, the reason that I propose is a "positive charge +q distributed in $V_Q$" with density $\rho^+=+q/V_Q$ C/cm$^3$. This is a fixed charge that I will call hereafter "*positive counterpart*" (PC) of each conductive electron in the CB. This PC of a carrier could keep trapped its cloud of charge $-q$ with a uniform density $\rho^-=-q/V_Q$ C/cm$^3$ everywhere in $V_Q$.

I am proposing that each carrier is a dipole of charges -*q* and +*q*, distributed evenly in $V_Q$, screening one to each other as best as possible. My proposal not the point-charge of -*q* coulombs "able to move" in $V_Q$. This blocking notion leads to a charged corpuscle that drifting under E=V/*d*, could travel the distance *d*. The dissipative current *i(t)=v(t)*/R that I am looking for, cannot rely on this idea. Going back to my extended dipole for each carrier, its PC grants the uniform distribution of its movable cloud of charge $-q$ in $V_Q$ that, at low T, was trapped in a small region around a $Si^+$ cation ("local dipole" of spherical form, no carrier existed yet). Gaining energy, this cloud passed to occupy the volume $V_Q$, thus sensing the force between its own charge $-q$ and that of its PC (+*q*). This dipole in $V_Q$ is the *new carrier of energy that I propose* from the trend of charge to keep charge neutrality in solid matter.

How this dipole appears or why the electronic cloud of each carrier senses its PC in this form falls out of the scope of this paper. It could be a "remembrance" of outer electrons and $Si^+$ cations that were together in the past. The point is that this fixed PC would keep local neutrality of charge at the atomic level in the lattice and overall charge neutrality in $V_Q$ for an external observer. I will assume too that the electronic cloud of each conductive electron does not collapse with its PC for reasons like those preventing its collapse with a $Si^+$ cation, when it is the outer electron of a Si atom. A carrier will be an "extended dipole in $V_Q$" of charges $-q$ (mobile) and $+q$ (fixed) ready to react to any electric field E=*v(t)/d*.

Readers used to handle point-like electrons should consider their radial electric field. This field emulates its PC of +*q* C distributed in a sphere at any distance from its point-charge $-q$. Looking at this electron at 1 km on my left, I would see negative charge from its electric field "leaving my eyes" but looking at my right, I would see positive charge from the same electric field "pricking my eyes". Near this electron, its PC lying at km or meters away does not matter too much. What matters is that it exists. Since dipoles of charge make equal or more sense than monopoles, let me continue with my PC of a carrier with density $\rho^+=q/V_Q$ C/cm$^3$ fixed to the atomic lattice, screened as best as possible by its movable cloud of $\rho^-=-q/V_Q$ C/cm$^3$.

Because this PC would be a "fine-grain" density of charge at the atomic level, each atom of the lattice should exhibit a tiny, positive charge to bind the mobile cloud. Thermal activity that once separated this cloud from a $Si^+$ cation that it could abandon gaining thermal energy, would try to separate this cloud from its PC. This means that this dipole of charges +*q* and -*q* at "mean distance" S(T) for each temperature T can store thermal energy. The electron with just the energy to reach the bottom of the CB would have the lowest S(T) possible $S_0$. Going up in energy, the "*conductive electrons of the CB*" would be extended dipoles of this type, with increasing S(T) values as their energy increases.

Since thermal activity would produce fluctuations of S(T), TEQ should define its mean value $S_{avg}(T)$ at a temperature T. This capacitor with charges -*q* and +*q* in its "plates" at distance $S_{avg}(T)$ on average, would store a mean energy of thermal origin $q^2/(2\langle C\rangle)$ in its mean capacitance $\langle C\rangle=C_f$ corresponding to the mean $S_{avg}(T)$. To store $U_f$=kT/2 J on average in this degree of freedom (DOF) that electrons (mobile clouds of carriers) can access, the mean capacitance $C_f$ of each carrier would be:

$$U_f=\frac{q^2}{2C_f}=\frac{kT}{2}\Rightarrow C_f=\frac{q^2}{kT}=\frac{q}{V_T} \tag{4}$$

Regarding the voltage *v(t)* between terminals, each carrier in the volume $V_Q$ of a resistor would react as a small capacitor of $C_f=q/V_T$ farads. Rapid fluctuations of S(T) due to thermal exchanges of energy always present, would set the "quiescent point" of this capacitance, defining its mean value $C_f$ to sense any *v(t)* like the dc one V=R×$I_C$. Let me recall now that Eq. (1) required that the energy left in $C_{mat}$ by each fluctuation was dissipated before the arrival of the next one. Knowing why λ does not depend on $C_{mat}$ in Eq. (1), I can say that no matter its value, capacitance is needed to collect packets of energy giving rise to fluctuations, and $C_f$ could play this role equally well. Moreover: $C_f$ *must be* this "collector of energy" because in TE, λ fluctuations per second entering packets of $U_f=q^2/(2C_{mat})$ J each, must sustain in time λ chances per second to dissipate the mean energy $U_f$=kT/2, see Eqs. (3) and (4). Equating λ×kT/2 to $\lambda\times q^2/(2C_{mat})$ one gets $C_{mat}=q^2/kT=C_f$.

Leaving shortly aside how fluctuations store energy in $C_f$, let me consider how to remove it quickly by the action of all the carriers of the resistor setting its conductance G=1/R. A fast way for a carrier to help with this task would be to convert the $U_f$ loaded in its $C_f$ into an energy ready to generate heat. To develop this idea, let me bias a resistor with a current $I_C$. This would add to its ac voltage of Johnson noise always present, a big dc term V=R×$I_C$, entailing a dc field $E_{Joul}$=V/*d* between its terminals. This $E_{Joul}$ would pull the cloud of negative charge towards the anode with a force equal to that pushing its PC towards the cathode. These two forces acting on each carrier would cancel one to each other without charges reaching any terminal. However, these two forces stretching the dipole of the carrier, would create a small strain of the atomic lattice that suggests how dissipation can occur. It would begin by $E_{Joul}$ displacing slightly the mobile cloud of this dipole from its mean positions in TE around each atom of the lattice.

Fig. 5-b shows this cloud pulled towards the anode, which in turn would displace towards the anode each atom from its mean position for V=0. This is what Fig. 5-b shows by its tilted bonds representing the strain of the lattice storing elastic energy

coming from $V=R\times I_C$. This energy that $C_f$ would store as elastic one by increasing S(T), would remain in the lattice while the cloud pulling the lattice and the voltage V, both exist. By this process, carriers would convert electrical energy set in $C_f$ by V into elastic energy $U_{Joul}$ stored in the lattice. Note that for $I_C=0$, the Johnson noise of the resistor would take the role of V to sustain this conversion process since Johnson noise needs an incessant dissipation of fluctuations [1] to exist in TE.

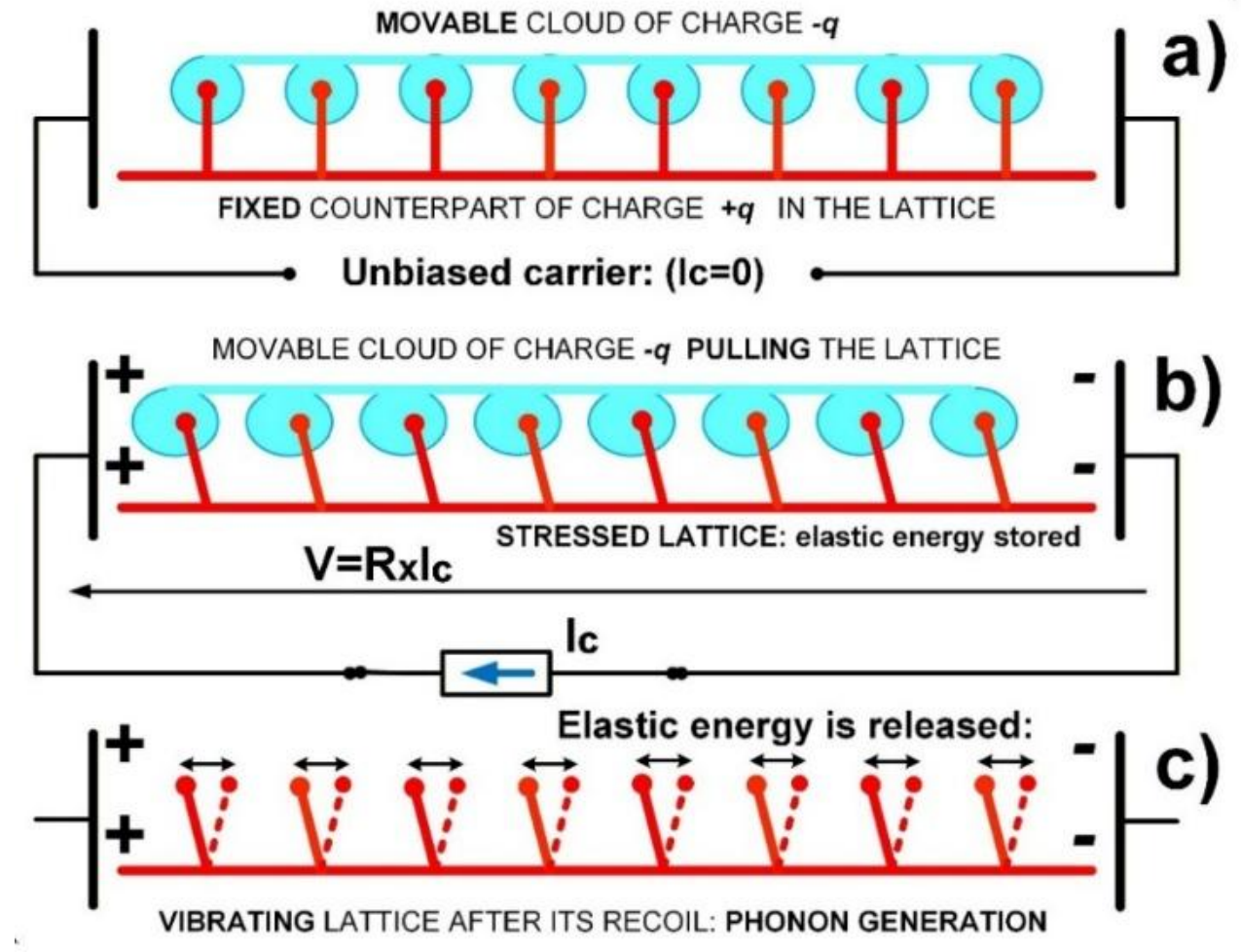

**Fig. 5** Dipolar structure of a conductive electron in the CB (a carrier): **a)** In thermal equilibrium (thus unbiased), **b)** under a bias current $I_C$, and **c)** vibrating lattice left by a biased carrier that disappears from the CB (see the text).

Since storing $U_{Joul}$ in this way deforms a dipole of charge between terminals, the capacitive current of each carrier to go from the situation of Fig. 5-a in TE to that of Fig. 5-b under $v(t)$=V, would come from the external generator of $I_C$. In TE however, such dc current $I_C$ would be replaced by the random current coming from the fluctuations that generate its Johnson noise $v(t)$. All in all, thermal activity plus the external generator, should deliver to the resistor the energy kT/2 plus $U_{Joul}$ (if $I_C\neq 0$) that each carrier would store as elastic energy ready to produce heat in $V_Q$. The field $E_{Joul}$ acting on $C_f$, whose plates of charges $-q$ and $+q$ would be at the mean S(T) set by Eq. (4), (small ac energy) would increase this S(T) to load also the usually higher energy $U_{Joul}$. Hence, each carrier would store ac energy of thermal origin (kT/2 on average) and dc energy $U_{Joul}$ due to $V\neq 0$, mimicking the hold capacitor of a S&H amplifier storing a small ac voltage (noise) superimposed to the dc one (signal) that it has sampled and that it holds for further processing. Thus, *carriers in resistors would carry energy* rather than charge.

Because electrons in the CB have access to several DOFs, the mobile cloud of each carrier could leave this DOF where it forms a carrier, to access other DOFs like to be again the outer electron of a donor atom or the electron filling a surface state within the 2TD. In both cases, we would say "the electron of the carrier has left the CB to be trapped again" and this leads to consider $\tau_{CB}$, its lifetime in the CB. Then, let me study what occurs when a carrier loaded with $U_{Joul}$, leaves the CB to be trapped by a $Si^+$ cation for example. I mean when such a carrier of lifetime $\tau_{CB}$ dies in this way, and its cloud of charge extended between terminals in $V_Q$ disappears. This means that its pulling action on the atoms of the lattice ceases suddenly. This would release the strained lattice, which would start to vibrate as a spring-loaded lever does if you release it suddenly. The elastic energy $U_{Joul}$ of this carrier just dead, would produce vibrations of the lattice (phonons propagating in this periodic medium), which would be the heat that we assign to $I_C$ by Joule effect

Dissipating the energy of a carrier when its mobile cloud no longer is between terminals avoids displacement of charge that would vary its Johnson noise. This grants the totally different roles that our model for EN assigns to each type of current that it uses. To account for $P_{DC}=R\times(I_C)^2$ watts that are dissipated in a resistor by its bias current $I_C$, the mean number of carriers that should "die" each second is $P_{DC}/U_{Joul}=\lambda$. Since each time a carrier dies its elastic energy is released as heat, Eq. (2) states that the rate of carriers disappearing from the CB will be λ, no matter the $I_C$ or the voltage $V=R\times I_C$ between terminals of the resistor. In my model, a bias current would not change the rate of fluctuations of a resistor (if its T does not vary noticeably) but it would not exist without them (see below). This shows the link between the three types of current of my model for EN: the instantaneous displacements of electrons for fluctuations, the subsequent recoils of charge in opposed sense (Nyquist noise) and the dissipative currents using carriers of energy.

## V. Revisiting Joule effect to test the quantum model for Johnson noise and dissipative current.

Despite the low value $C_f$=6,2 attofarad at room T, this tiny capacitance per carrier gives relevant dissipations in resistors of macroscopic size. Regarding my cubic resistor of R=50Ω and $f_c$=32GHz, and assuming full ionization at room T for its low $N_d$, it would contain $n\approx N_d\times V_Q=10^{12}$ carriers. From its $\tau_d$=5 ps (or its R=50Ω shunting its $C_{mat}$=0,1pF), the power that it would dissipate in TE at room T is $P_{TE}=kT/\tau_d$=820pW. Biased by $I_C$=4 mA its V=0.2 volts between terminals give $P_{Dis}=V^2/R$=800μW, that roughly is one million times $P_{TE}$. Thus, each carrier would load from V the energy $U_{Joul}=1{,}24\times 10^{-19}$ J of Eq. (3) that would be $\approx 10^6$ times higher than $U_f$. Since its mean rate of fluctuations $\lambda=6{,}43\times 10^{15}$ $s^{-1}$ must be the rate of carriers dying each second in its $V_Q$, the lifetime of its carriers would be: $\tau_{CB}=n/\lambda$=156μs.

Each carrier of my cubic resistor would die 6410 times per second on average no matter if it dissipates $P_{TE}$=820pW in TE or $P_{Dis}$=800μW under $I_C$=4mA. I am assuming that $P_{Dis}$ does not modify noticeably its temperature of TE. Since $\tau_{CB}$=154μs is $3\times 10^8$ times $\tau_d$=5ps, its carriers are "plenty of time" to charge their $C_f$ with V volts and I will not consider here this subject to keep short this paper. Doubling its length to get a 100Ω resistor made from two cubic ones in series and for the same $I_C$, this longer device would dissipate twice the power of my cubic one. Due to its R=100Ω=2×50Ω, the rate $\lambda^*$ of this longer device would be half the rate of my cubic one, $\lambda^*=\lambda/2$ see Eq. (1).

This 100Ω resistor and mine of 50Ω, both having the same material, would dissipate the same power $P_{TE}=kT/\tau_d$ at room T. Since the voltage between terminals of the longer device (0,4 volts for $I_C$=4mA) doubles that of my cubic one, the energy that its carriers would load is four times higher. This agrees with its rate $\lambda^*=\lambda/2$ of carriers dying each second to dissipate twice the power of mine $P_{dis}^*=\lambda^*\times 4U_{Joul}=2\times 800\mu W$. Since the number of carriers in this resistor doubles that of my cubic one, but its rate of carriers dying is $\lambda^*=\lambda/2$, their lifetime is $\tau_{CB}^*=n^*/\lambda^*=4\tau_{CB}$, thus $\tau_{CB}^*=616\mu s$. Doubling the length of the resistor the lifetime of its carriers becomes four times longer [8].

Connecting two of my cubic resistors in parallel would give a wider one of R=25Ω dissipating $P_{TE}=kT/\tau_d$=820pW in TE at room T (same material). For $I_C$=4mA, this 25Ω resistor would dissipate half the power of my cubic one because halving the resistance, the rate $\lambda^*$ of the wider device is twice the rate λ of mine ($\lambda^*=2\lambda$). Since the voltage between terminals for $I_C$=4mA is halved (V=0,1 volts) the energy loaded in each carrier would be four times lower. Because the rate $\lambda^*$ of carriers dying per second in this wider resistor is $\lambda^*=2\lambda$, its dissipation for $I_C$=4mA $P_{dis}^*=\lambda^*\times U_{Joul}/4=400\mu W$ is right. Since this 25Ω resistor has twice the carriers of mine and its rate of deaths $\lambda^*=2\lambda$ doubles that of mine, their lifetime $\tau_{CB}^*=n^*/\lambda^*$ is equal to $\tau_{CB}=n/\lambda=154\mu s$ in my cubic one. Keeping the length of the resistor, the lifetime of its carriers does not change.

Let me resume my reasoning on $C_f$ as the "antenna" picking up thermal energy that I left aside in the second paragraph under Eq. (4). I mean to consider "*how a fluctuation sets energy in $C_f$*". Due to the structure of each carrier (a fixed PC of $\rho^+=+q/V_Q$ C/cm$^3$ that is screened everywhere by its cloud of $\rho_{cloud}=-q/V_Q$ C/cm$^3$), this mobile cloud is an electron that absorbing a packet of thermal energy could appear instantly in terminal B. At first sight, this translocation would leave unscreened in $V_Q$ its PC of density $\rho^+=+q/V_Q$ C/cm$^3$, as Fig. (6) shows. The built-in voltage Δψ of this dipolar space charge region supposedly left in $V_Q$ can be found from $E_{peak}=-q/(\varepsilon\times A_p)$, the field aside the plate B where this negative charge $-q$ C has just arrived. I refer to the electric field on the inner surface of terminal B of area $A_p$, that would vary linearly up to zero as we move towards terminal A in Fig. (6). This forms a triangle of area $\Delta\psi=(E_{peak}\times d)/2$ that is:

$$\Delta\Psi = V_A - V_B = -\frac{-q}{\varepsilon\times A_p}\times\frac{d}{2} = \frac{1}{2}\frac{q}{\frac{\varepsilon\times A_p}{d}} = \frac{1}{2}\frac{q}{C_{mat}} \qquad (5)$$

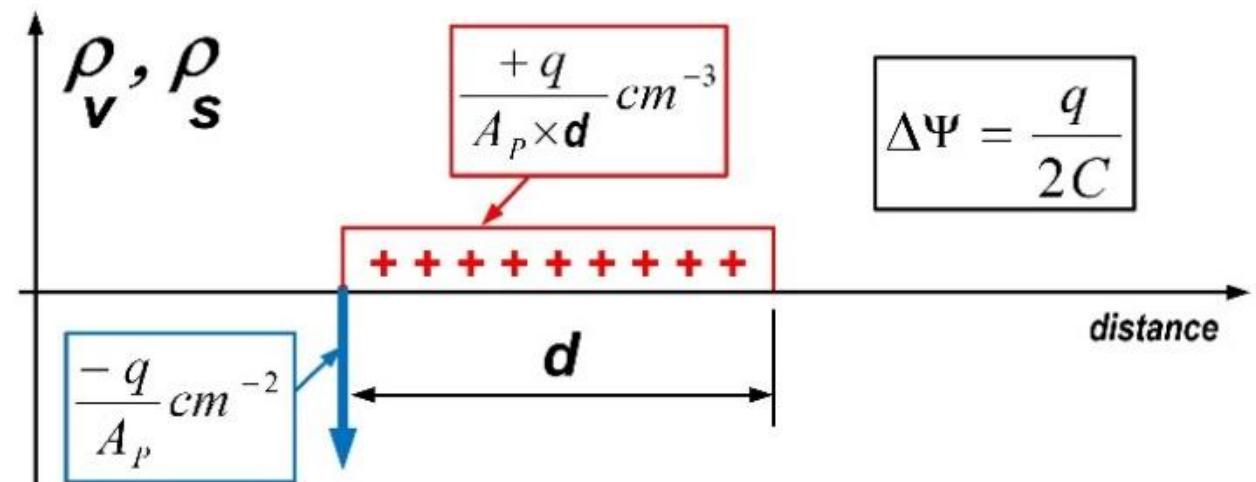

**Fig. 6** Charge densities if the cloud of a carrier appearing suddenly in one terminal, could leave its PC unscreened in $V_Q$.

Eq. (5) seems the solution of Poisson's equation to get the energy $U_f=q^2/(2C_{mat})$ used in Eq. (1). However, the voltage step $\Delta\psi=(V_A-V_B)$ that appears between terminals in this case *is not* $\Delta V=q/C_{mat}$ volts that I have used for fluctuations where a single electron was translocated between terminals. This step only is ΔV/2 to warn that "an electron has arrived in terminal B without an electron leaving simultaneously terminal A". This *prevents us from contravening the current continuity* that the capacitive coupling of these terminals demands. Hence, if the electronic cloud of a carrier appears suddenly in one terminal of a resistor, the electric field of its inner surface $E_{peak}=-q/(\varepsilon\times A_p)$ will set a field $E_{peak}=+q/(\varepsilon\times A_p)$ simultaneously in the inner surface of the opposed terminal at distance *d*.

Thus, if the mobile cloud of a carrier disappears from $V_Q$ to appear located on one terminal as a charge density $-q/A_p$ C/cm$^2$, an opposed density $+q/A_p$ C/cm$^2$ will appear simultaneously in the other terminal. I would say that the "unscreened PC" of the carrier in $V_Q$ that Fig. 6 shows, becomes instantly a *charge +q* on the surface of the terminal in front. This fact doubles Δψ in Eq. (5) and solves the problem giving $\Delta\psi=q/C_{mat}$, thus $\Delta V=q/C_f$ as it must be. This suggests that in devices of two terminals like a resistor, the PC that I have proposed for a carrier could be a form to imagine how the terminals of this 2TD interact with the dipolar structure of its carriers. Note the need to keep current continuity to obtain cogent results.

All in all, when a carrier dies because its electronic cloud appears in one terminal, a charge $+q$ appears in the other at the same instant. In my first model for EN, this translocation of an electron between terminals was a fluctuation storing capacitive energy $U_f$ in the $C_{mat}$ of of the resistor. In my next model for EN, this fluctuation of energy is stored in the $C_f$ of each carrier, not in the macroscopic $C_{mat}$. I have given reasons for it. Since this fluctuation removes from $V_Q$ its electronic cloud, the elastic energy that this carrier could have before dying in this way would pass to the lattice as heat. Given the huge number of carriers in macroscopic devices, this type of fluctuation would be overwhelming. Although I have kept the macroscopic $C_{mat}$ in Eq. (5) to remember the time when I solved Poisson equation in AlGaAs devices, I have shown clearly that $C_f$ is the actual "collector" of thermal energy in resistors.

The idea of a macroscopic antenna ($C_{mat}$) absorbing packets of thermal energy $U_f$ that a resistor would dissipate to show its Johnson noise, has led to the notion of a resistor that does this absorption by the myriad of electronic antennas it would have between its terminals. I refer to my proposal for each carrier as an extended dipole able to exchange energy. This myriad of antennas absorbing randomly packets of energy $U_{Joule}$ if $I_C\neq 0$ plus energy $U_f$ on average, at the huge rates λ of Eq. (1) would produce its Johnson noise while emulating quite "perfectly" a dissipative current that hardly would exist: "dc" current. I mean "constant" current emulated by the random dissipation at a huge mean rate λ of packets of energy $U_{Joule}$.

Eq. (5) and Fig. 6 remind the capacitive coupling between terminals that entails the current continuity that we must keep in electrical measurements. To contend that an electron has arrived in one terminal of the resistor, we must consider that an electron has left simultaneously the other. To apply this notion to the extended dipole of a carrier we must consider that its electronic cloud is coupled by $C_f$ to its PC of density $+q/V_Q$ C/cm$^3$ that it tries to screen as best as possible. Thus, if the electron of a carrier arrives suddenly in terminal A (plate A of $C_f$), this is so because an electron leaves its terminal B (plate B of $C_f$) simultaneously and the energy of this event will be $q^2/(2C_f)$. Since $C_f$ couples the cloud and the PC of each carrier, when this cloud appears on one terminal, the charge $+q$ of its PC appears simultaneously in the other. A voltmeter measuring these simultaneous events will give the voltage $\Delta V=q/C_f$, which is the effect of this impulsive current that we cannot measure directly. Some of these ideas already were in [8].

Concerning a carrier that dies because its electronic cloud is trapped by a $Si^+$ cation, or because it falls into a surface state somewhere between terminals, none of these capture events should give a fluctuation nor its shot noise between terminals because no electron would arrive in one of them. The elastic energy that the dying carrier could have would pass to the lattice because any of these capture events removes its cloud of charge between terminals. A subsequent "emission" to the CB of this electronic cloud thus trapped, would not give shot noise either but only a change in "carrier number" whose possible resistance noise falls out the scope of this work.

This would complete this quantum model for EN where the fluctuation more frequent in resistors would be measured as the translocation of an electron between terminals. I refer to the cloud of charge of a carrier appearing instantly in one of its terminals and its PC appearing simultaneously in the other. This event should give a voltage step $q/C_f=kT/q$ volts called thermal voltage $V_T=kT/q$. This removal from $V_Q$ of the cloud of charge of the carrier would release as heat the elastic energy that this carrier had when this fluctuation killed it: kT/2 J on average plus $U_{Joule}$ if $I_C$ existed. Dissipations like this one would relax the voltage between terminals with the time constant $\tau_d$ that gives rise to $i_d(t)$ removing the charge unbalance and to $i_c(t)$ accounting for this dissipation. In this way, the PC of each cloud that formed a carrier would be available to form a new carrier that loading thermal energy and $U_{Joule}$ if needed, would be ready to produce a new fluctuation. These fluctuations taking place with mean rate $\lambda$, would sustain in time the dissipation that Johnson noise needs to exist, as well as the added dissipation that Joule effect would demand in a resistor biased by $I_C\neq 0$.

## VI. A DEMANDING TEST OF THIS MODEL: PHASE NOISE.

In the year 2008, I thought that a way to check the reach of my first quantum model for EN would be to predict the phase noise (PN) of an electronic oscillator (EO). I mean the random phase modulation of its output signal expected to be a pure sinusoid of voltage with precise frequency $f_o$ called "carrier" that an EO never gives. What it gives is a phase-modulated carrier of instantaneous frequency varying randomly around $f_o$. This instability is seen as PN coming from the EN of its LCR resonator plus EN of its feedback electronics.

Without a quantum model for EN, it is hard to show how EN translates into PN. To deal with this subject, let me focus on an EO formed by the LCR resonator of an L-C tank, properly connected to an amplifier to sustain oscillations in the electronic loop thus formed. Let me say that an inductance L in parallel with a capacitance C form the "L-C tank" where electric and magnetic energies exchanged in time, give the output signal sustained at $f\approx f_o$ in this resonator by its feedback electronics.

Let me note that to have a pure sinusoid of voltage in C or L at frequency $f_o$ in this EO, the magnetic energy in L and the electrical one in C will fluctuate at $2f_o$. Since L and C both have voltage and current *in quadrature*, they would follow perfectly such fluctuations due to the lack of dissipation in this simplified circuit. I refer to the lack of *current in phase* with the carrier, a situation that changes when we consider losses in this circuit. To account for losses in L (radiative, magnetic and those by Joule effect) and in C (mainly electrical losses by its shunting conductance) we must give room for dissipative current by a resistance R sensing the voltage of C. This 2TD with three elements (L, C and R) is what I have called "LCR resonator".

In the loop of the EO, this resonator will sense the loading effect of its electronics decreasing R down to $R^*$. Thus loaded, this resonator will undergo $\lambda=2kT/(q^2R^*)$ fluctuations/s of its capacitive energy as translocations between terminals of single electrons that it will dissipate by "all its losses" including those due to the noise factor F of its electronics making F times larger its previous rate $\lambda$. All the above would generate its voltage noise of Lorentzian spectrum centered at $f_o$ (thus a pass-band one). To predict the PN of the output of this EO let us think of its output accumulating phase in time under null dissipation ($R^*\rightarrow\infty$). This gives a straight line with a slope $2\pi f_o$ rad/s that acquires a kind of small random walk around this slope due to the $F\lambda$ fluctuations in its resonator. Considering the *phase shifts* of the $F\lambda$ responses h(t) per second in the carrier, *each weighted by its effect depending on its position within the period* $T_o=1/f_o$, we obtained the PN of this EO coming from its losses and those added by its electronics [9, 10].

Our results confirmed exactly the semi empirical formula of Leeson [11], thus giving it a theoretical support similar to that also given to works like [12] on the similarity between the PN of EOs and the linewidth of lasers. I refer to this quantum model for EN showing that *spontaneous translocations* of electrons between terminals of an L-C tank is the counterpart of the *spontaneous emissions* of photons in the Fabry-Perot resonator of a laser. About the suitability of impulsive noise like ours to predict phase noise of EOs, let me highlight the good agreement found in [13] between the PN measured in their EOs and its simulation by impulsive pulses of current in suitable capacitors of their circuits. This shows the interest of impulsive noise like ours that would be a "fine grain" model using monoelectronic fluctuations instead "multi-electronic" ones like the impulsive current of 179 fC ($\approx 1.1\times 10^6$ electrons) for a capacitor in [13].

Although Pauli exclusion principle prevents two electrons from behaving alike, this crowd of electrons crossing at the same instant such capacitor, likely reflects the fact that the shot noise density of $\lambda$ electrons of charge $q$ C passing randomly between two terminals each second and that of $\lambda/n^2$ fat electrons of charge $nq$ C each, would be similar. In any case, a good merit of [13] is to show that using impulsive currents in capacitors gives simulations in excellent agreement with measurements of PN. This agrees with the notion of fluctuation as an impulsive current that "already is gone" when you notice its effect. This is why an automatic level control (ALC) nothing can do in an EO to prevent or reduce its PN close to the carrier that comes from fluctuations [9, 10]. By contrast, it affects the pedestal of PN far from the carrier that comes from dissipative currents in phase with the carrier being generated.

## CONCLUSIONS

We present a model for electronic noise within the quantum framework of Callen and Welton that explains the Johnson noise of resistors by instantaneous displacements of single electrons between their terminals. The subsequent relaxations of the charge unbalances thus created (fluctuations) entail two currents (capacitive and dissipative) of opposed sign, cancelling mutually at each instant of time as they generate Johnson noise. Since the shot noise of the capacitive currents accounts for all the observed noise, their opposed currents that are dissipative should not produce noise voltage, as it is observed over and over in resistors biased by dc currents.

Because electrical current and electronic noise go together and dissipative current as a net flow of charge is a widely used notion for dissipative current, we present a disruptive model for electrical current that dissipate energy without affecting those displacement currents already present in resistors. From charge neutrality, current continuity and using carriers of energy, this model backs the quantum model for EN of this work.

Charge relaxations giving rise to Johnson noise suggest to review our notions on Nyquist noise, its spectral density and its conversion into voltage out of the device where it is generated. The capacitive *i-v* converter of each resistor (i. e. capacitance between its terminals) is much faster than external converters connected to extract its Nyquist noise from the capacitive one where this random current already has caused its effect: the voltage that we measure called Johnson noise.

Concerning the notion of dc current as “constant” current, its impulsive origin driven by random fluctuations of energy at huge rate $\lambda$ on average, is worth noting. I mean its fully rectified appearance coming from fully random “ac” events: fluctuations of thermal origin between the terminals of the resistor.

New notions derived from these notions could help to fight against myths in this field like “ohmic” resistances considered noisy while dynamic ones are taken as noiseless. Simulation programs like PSPICE could benefit from this model for their simulations under null net current coming from two opposed ones in TE for example.